\documentclass[pra,showpacs,twocolumn,superscriptaddress,amsmath,amssymb,longbibliography]{revtex4-1}
\pdfoutput=1
\usepackage{bm}
\usepackage{graphicx}
\usepackage{color}
\usepackage{dcolumn,amsmath}
\usepackage[normalem]{ulem}

\begin{document}

\title{Universality and Chaoticity in Ultracold K+KRb Chemical Reactions}

\author{J. F. E. Croft}
\affiliation{Department of Chemistry, University of Nevada, Las Vegas, Nevada 89154, USA}
\author{C. Makrides}
\affiliation{Department of Physics, Temple University, Philadelphia, Pennsylvania 19122, USA}
\author{M. Li}
\affiliation{Department of Physics, Temple University, Philadelphia, Pennsylvania 19122, USA}
\author{A. Petrov}
\affiliation{Department of Physics, Temple University, Philadelphia, Pennsylvania 19122, USA}
\affiliation{NRC ``Kurchatov Institute'' PNPI, Gatchina, Leningrad district 188300, Russia}
\affiliation{Division of Quantum Mechanics, St.Petersburg State University, 7/9 Universitetskaya nab., St. Petersburg, 199034 Russia}
\author{B. K. Kendrick}
\affiliation{Theoretical Division (T-1, MS B221), Los Alamos National Laboratory, Los Alamos,
New Mexico 87545, USA}
\author{N. Balakrishnan}
\email{Correspondence should be addressed to NB (naduvala@unlv.nevada.edu).}
\affiliation{Department of Chemistry, University of Nevada, Las Vegas, Nevada 89154, USA}
\author{S. Kotochigova}
\affiliation{Department of Physics, Temple University, Philadelphia, Pennsylvania 19122, USA}

\begin{abstract}
{\bf A fundamental question in the study of chemical reactions is how
reactions proceed at a collision energy close to absolute zero.
This question is no longer hypothetical: quantum degenerate gases of atoms and
molecules can now be created at temperatures lower than a few tens of
nanoKelvin.
Here we consider the benchmark ultracold reaction between, the most-celebrated
ultracold molecule, KRb and K.  For the first time we map out
an accurate {\it ab~initio} \/ground state potential energy surface of the
K$_2$Rb complex in full dimensionality and  report numerically exact
quantum-mechanical reaction dynamics.
The distribution of rotationally resolved rates is shown to be Poissonian.
An analysis of the hyperspherical adiabatic potential curves
explains this statistical character revealing a chaotic  distribution for the
short-range collision complex that plays a key role in governing the reaction outcome.
}
\end{abstract}
\maketitle

\newpage

The ability to prepare reactants and control products on demand with quantum
state precision is chemistry's holy grail \cite{Science2010,
bell.softley:ultracold,hutson:ultracold,carr.demille.ea:cold,krems:cold,
bethlem.meijer:production,dulieu.krems.ea:physics,balakrishnan:perspective,krems:viewpoint,
Knoop2010}.
Ultracold chemistry is a new and rapidly progressing field where reactants are
prepared in a single quantum state which holds out this promise~\cite{Science2010,
rellergert.sullivan.ea:evidence,knoop.ferlaino.ea:magnetically,tscherbul.krems:tuning}.
In a pioneering experiment, groups at JILA were able to produce an ultracold gas
of $^{40}$K$^{87}$Rb molecules at nano Kelvin temperature where even the
nuclear spins were oriented~\cite{Science08,Science2010}.  The exothermic
reaction rate coefficients of KRb+KRb and KRb+K were also measured. In
addition, by merely flipping a nuclear spin the KRb+KRb reaction could be turned on and
off. This is a perfect illustration of an on demand reaction demonstrating the
control that is attainable in ultracold gases.  Such control also promises to
answer open questions such as the role of geometric phases, non-adiabatic
\cite{garand.zhou.ea:nonadiabatic} and Van der Waals entrance-channel
effects~\cite{skouteris.manolopoulos.ea:van}, and the effect of external fields
on the product distribution in chemical reactivity. Detection of complex
molecules with single quantum state precision has been shown to be feasible in
a hotter 1 K environment~\cite{patterson.schnell.ea:enantiomer-specific,
spaun.changala.ea:continuous}.

However, theoretical calculations of reaction dynamics for such systems pose
a daunting computational challenge.
Ultracold collisions are sensitive to details of the potential and
require accurate electronic potential calculations. In addition many of the
systems of current interest are heavy with deep potentials, meaning that
scattering calculations need to include many channels with computational costs
scaling with the cube of the number of channels. It is for these reasons that
while the pioneering experiments \cite{Science08,Science2010}, on KRb reactions,
were performed over six years ago no accompanying scattering calculations have
been performed, until now.
Such calculations are needed to provide key predictions for
state-to-state rate coefficients, which await laboratory measurement.

The reaction of KRb with K is relatively fast with
a rate coefficient of the order of $10^{-10}$ cm$^3$/s which agrees
fairly-well with predictions of a universal model (UM),
based on quantum defect theory \cite{Julienne2010,Kotochigova2010}. This
suggests that most molecules undergo atom-exchange and only a small
percentage undergo an elastic collision.
A statistical approach has been used to analyze the effect of an
external electric field on the rotational product distribution in the
K+Rb$_2$ and KRb+KRb reactions~\cite{Maykel2014}.
However in order to understand the chemistry of such reactions one needs to
go beyond these approaches and examine reaction rates at a
state-to-state level.


The reactive scattering of complex atoms and molecules has always been
intricately intertwined with classical and quantum chaos.
In particular, classical trajectory simulations of reactive scattering show
self-similar behavior of dwell times between entering and leaving the central part
of the high-dimensional potential~\cite{Rankin1971,Kovacs1995,Barr2009}.
Recently there has been interest in the chaotic character of ultracold
inelastic collisions~\cite{frisch.mark.ea:quantum,maier.kadau.ea:emergence,frye.morita.ea:approach,green.vaillant.ea:quantum}.
Atom+dimer collisions involving three identical alkali atoms at ultracold energies have
been shown to be classically chaotic~\cite{Croft2014}.
This chaotic character has been taken as the starting point for work examining
statistical aspects of non-reactive ultracold alkali-metal dimer collisions~\cite{Mayle2012,Mayle2013}
and ultracold resonance reactions~\cite{flambaum.ginges:resonance}. Such works
suggest an approach to tackling ultracold reactions involving heavy alkali
species avoiding the prohibitive computational cost of numerically exact
calculations.

We report an explicit quantum mechanical study
of the ultracold reaction between a K atom and a KRb molecule.
To this end, we compute an
accurate {\it ab~initio}\/ ground-state potential energy surface (PES) of the KRbK complex
in full dimensionality, taking special care to accurately describe the long-range forces important in ultracold collisions. 
The total rate is shown to be universal in character -– validating the use of
simple universal models for other similar reactions.
On the other hand, the product rotational distribution is shown to be
statistical in character, which we attribute to the chaotic nature of the
reaction complex.

\vspace*{6mm}

\noindent

{\bf Results}\\

{\bf Potential Energy Surface Calculation}\\
The first theoretical studies of the electronic structure of three- and four-body alkali-metal systems
focused on homonuclear trimers~\cite{Soldan2007,Ernst2008,Simoni2009}.
Initial studies of heteronuclear alkali-metal trimer and tetramer potentials
principally located optimized geometries and dissociation
energies~\cite{JHutson2010,Cote2010,LopezDuran2015}.

In an alkali-metal  trimer the three valence electrons, one from each atom,
couple during the reaction and create two doublet and one quartet
adiabatic potential surfaces~\cite{JHutson2010}. The energetically lowest is a
doublet $^2A'$ potential. We have studied the reaction dynamics along this
potential surface and it is the focus of our electronic structure calculation.

Complete and accurate information on the lowest KRbK potential surfaces is unavailable.
Their computation requires substantial effort due to the complexity of the
multi-electron and open-shell reactants. In this work we perform a systematic
{\it ab~initio}\/ study  using the multi-reference configuration-interaction
(MRCI) method of the chemistry package MOLPRO~\cite{molpro}.
Details of this calculation can be found in Methods.

\begin{figure*}
\includegraphics[scale=0.37,trim=0 0 0 20,clip]{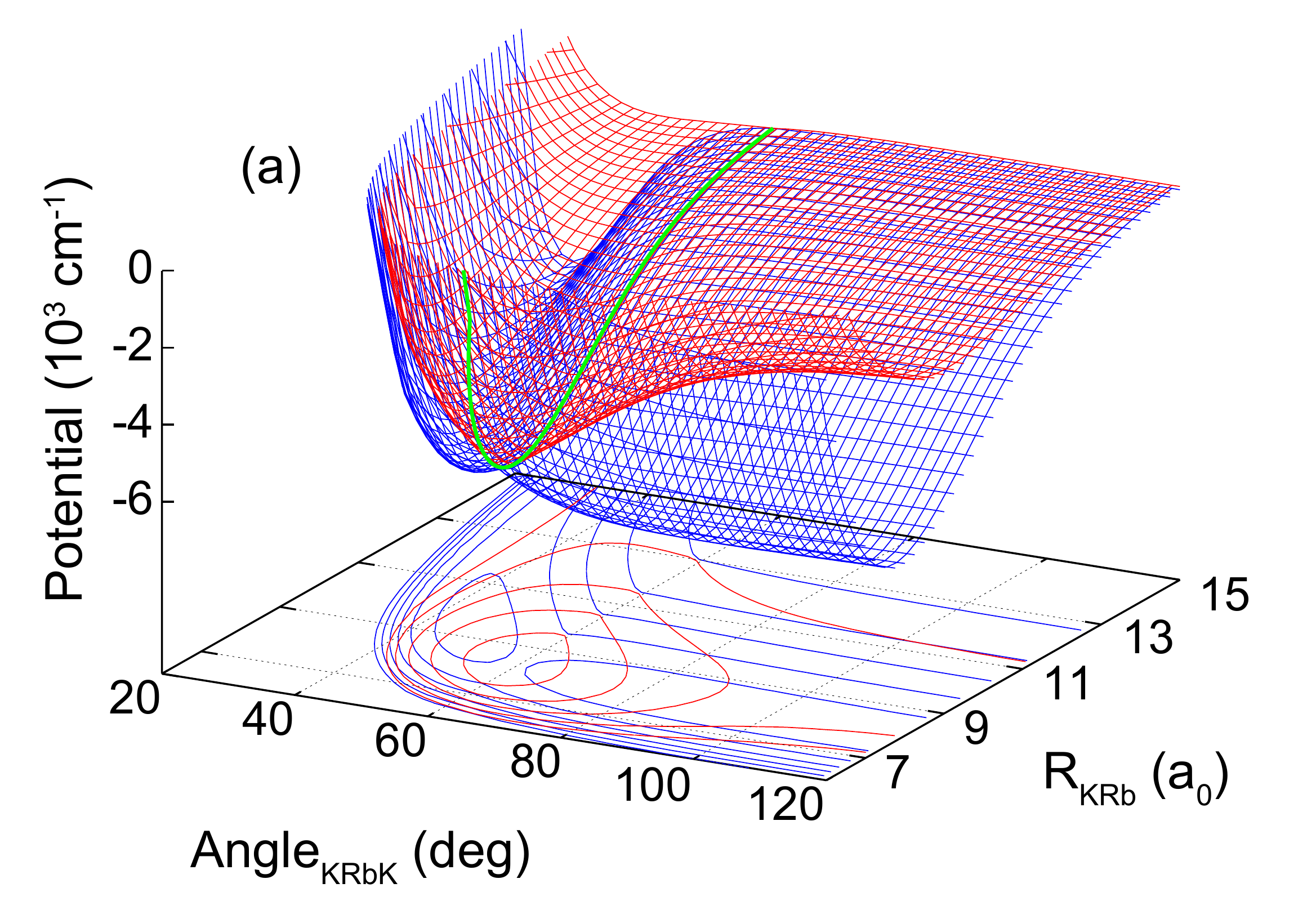}
\includegraphics[scale=0.25,trim=0 0 0 0,clip,angle=90]{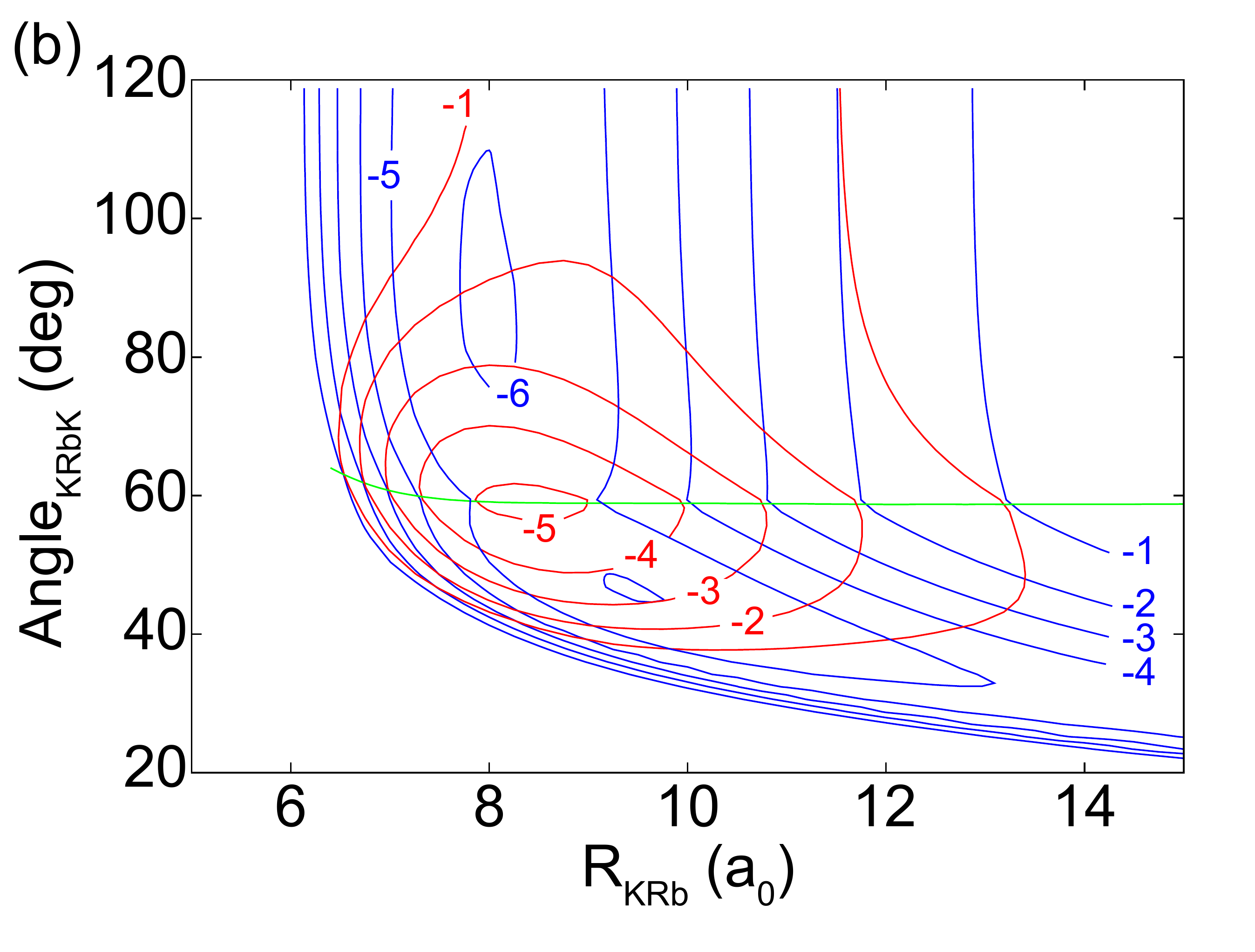}
\caption{{\textbf{$\vert$ K-KRb potential energy surface.}} Panel a: A two-dimensional cut through the energetically-lowest
$^2A'$ (blue curve) and $^2B'$ (red curve) adiabatic potential energy surfaces
of the KRbK trimer as a function of the K$-$Rb bond lengths and the angle
between K$-$Rb$-$K along the isosceles $C_{2v}$ geometry with
$R_{\rm K_1Rb} = R_{\rm K_2Rb}$. The base of the figure shows the corresponding
contour graph. The zero of energy corresponds to the energy of three
well-separated atoms. The unit of length is the Bohr radius $a_0=0.0529177$~nm
and that of energy is wavenumbers.
Panel (b) shows  a contour graph of the K$-$Rb$-$K potentials based on the data
in panel (a). Contour labels are in units of $10^3$ cm$^{-1}$. In both panels
the seam of conical intersections is shown by the green line.}
\label{KRbK}
\end{figure*}

Figure~\ref{KRbK}a shows a two-dimensional cut of the energetically-lowest
adiabatic potentials, $^2A'$ and $^2B_2'$, along the isosceles $C_{{\rm 2v}}$
geometry where $R_{\rm K(1)Rb}=R_{\rm K(2)Rb}$.
The reactant and product states are situated in the pairwise potential wells when either
$R_{\rm K(1)K(2)}$ or $R_{\rm KRb}$ is large.  We find that the lower
$^2A'$ potential has an absolute minimum
when $R_{\rm K(1)Rb}=11.10 a_0$,  $R_{\rm K(2)Rb}= 8.04 a_0$, and
$R_{\rm K(1)K(2)}= 7.64 a_0$.
The atomization energy, the minimum energy of three seperated atoms
measured from the potential minimum, is $V_{\rm A}$ = 6258 cm$^{-1}$.
The dissociation energy from the optimized geometry and the limit KRb + K is
$V_{\rm d1}=2079$ cm$^{-1}$, while that to the limit K$_2$+ Rb is
$V_{\rm d2}=1854$ cm$^{-1}$.

The accuracy of our {\it ab~initio}\/ trimer potential is tested by calculating
the {\it ab~initio} \/ dimer X$^1\Sigma^+$ and a$^3\Sigma^+$ potentials for
both KRb and K$_2$ at the same level of electronic-structure theory as the
trimer potential.
These potentials are compared with their corresponding spectroscopically-accurate
dimer potentials for KRb~\cite{Tiemann2007} and
K$_2$~\cite{Tiemann2008} in Methods.
As discussed in Methods comparison between our
theory and experiment shows an excellent agreement.  In addition, we use the dimer X$^1\Sigma^+$ and
a$^3\Sigma^+$ potentials to construct the lowest pairwise trimer potentials
for KRbK following the  dimer-in-molecule theory of Ref.~\cite{Ellison1963}.
Details of the analytical construction of these potentials are given in Methods.

\begin{figure*}
\includegraphics[scale=0.18,trim=10 -35 50 100,clip]{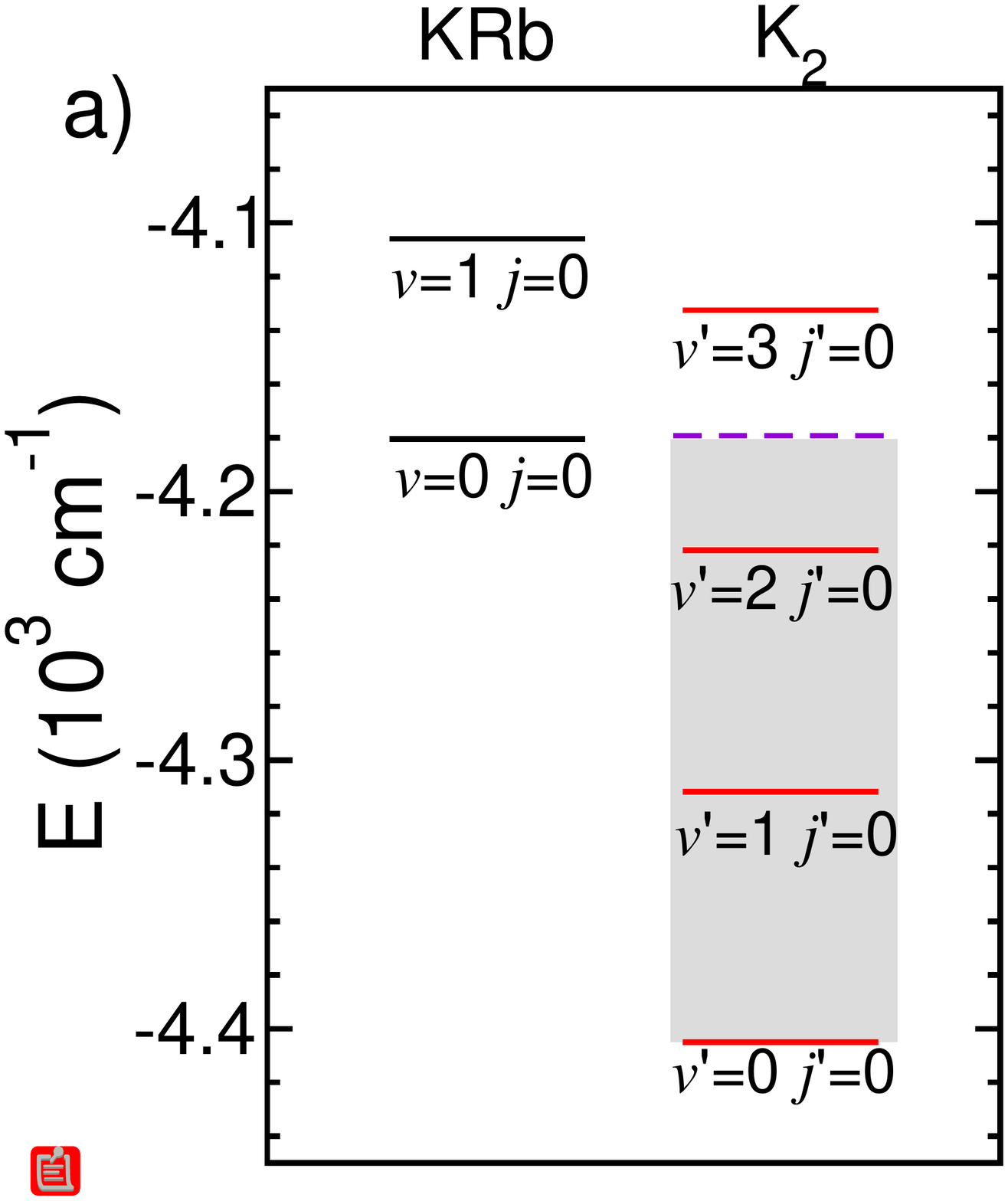}
\includegraphics[scale=0.21,trim=20 0 0 20,clip]{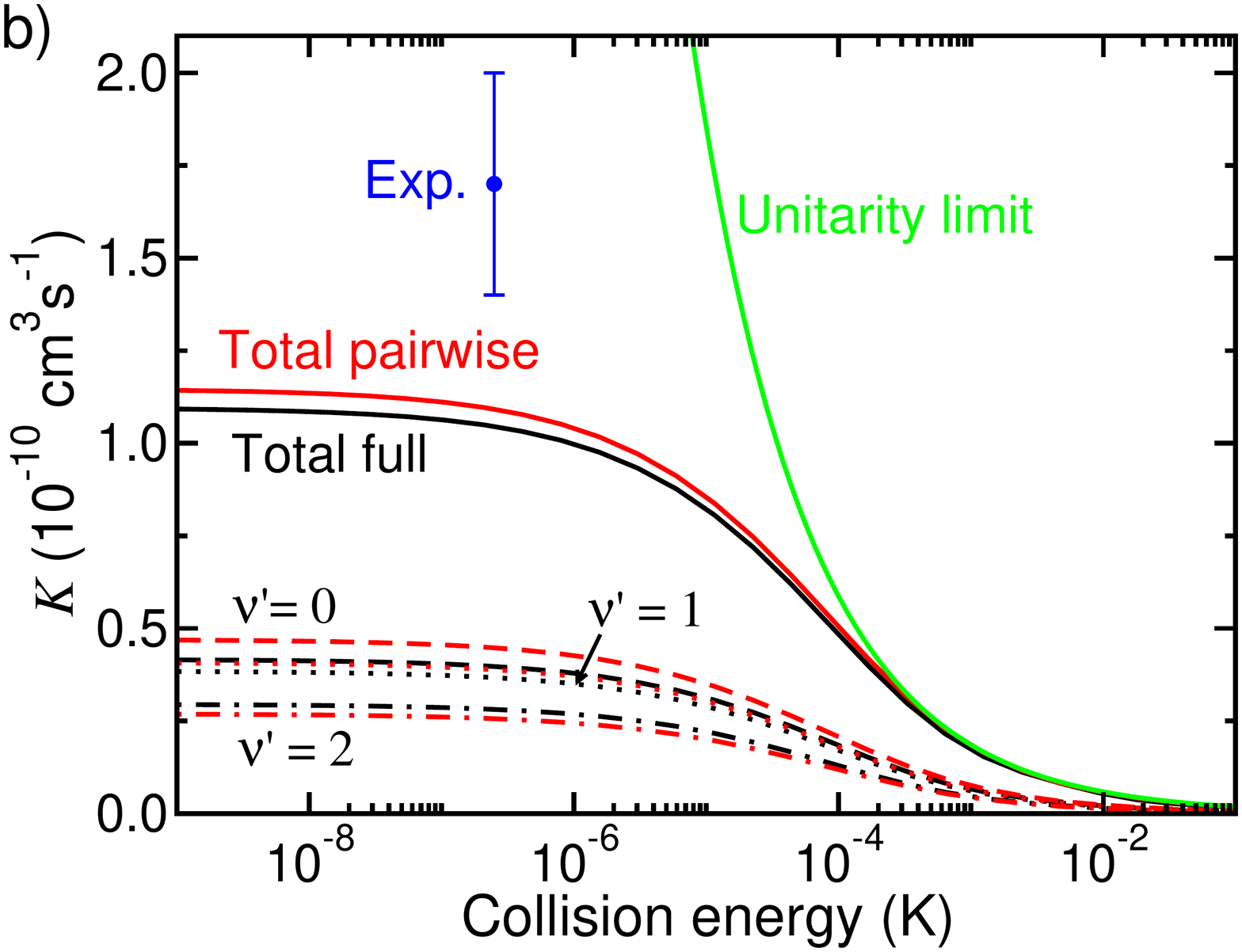}\includegraphics[scale=0.21,trim=20 0 0 20,clip]{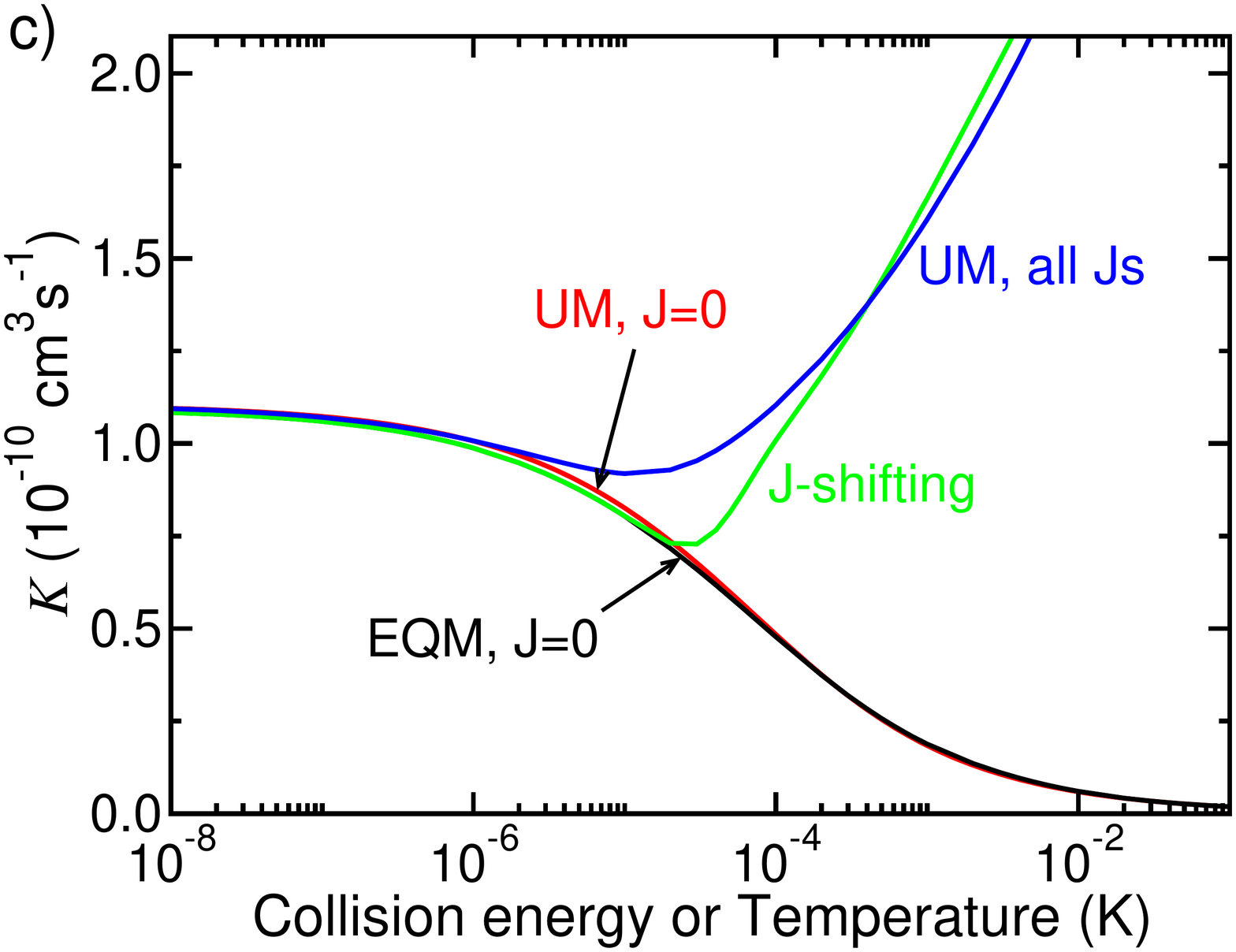}
\caption{{\textbf{$\vert$ Energetics and reaction rates.}} Panel a: Energetics of the KRb + K $\to$ K$_2$ + Rb reaction.
The $j=0$ vibrational levels of the KRb X$^1\Sigma^+$ potential and $j'=0$
vibrational levels of the K$_2$ X$^1\Sigma^+_g$ potential are shown by
black and red lines, respectively.
The gray shaded area indicates the closely-spaced energetically-allowed
rotational levels of K$_2$. The zero of energy is located at the dissociation
limit of KRb and K$_2$.
Panel b: Reaction rate coefficients from $J=0$ EQM calculations based on either
the full (black curves) or pairwise (red curves) potential as a function of
collision energy in units of the Boltzmann constant. The total and
vibrationally resolved reaction rate coefficients are shown. The green curve is
the $s$-wave unitary rate coefficient for atom-dimer scattering.
The closed circle with error bars  (one s.d.) corresponds to an experimental
measurement~\cite{Science2010} taken at a temperature of 250~nK.
Panel c: Total reaction rate coefficient (green curve) as a function of
temperature based on the $J$-shifting method and the $J=0$ EQM results for the
full trimer potential. The black curve repeats this latter curve from panel b
as a function of energy. The red curve shows the rate coefficient of the $s$-wave
universal model (UM) as a function of collision energy, while the blue curve
shows the UM data including all relevant partial waves as a function of temperature. }
\label{fig:rates}
\end{figure*}

\vspace*{4mm}
\noindent
{\bf Quantum Dynamics Calculations}\\
We present exact quantum-mechanical (EQM) calculations for the
KRb($^1\Sigma^+, v=0,j=0$) + K($^2$S) $\rightarrow$ K$_2(^1\Sigma_g^+, v',j'$) + Rb($^2$S)
chemical reaction, where $v,v'$ and $j,j'$ are vibrational and rotational
quantum numbers, respectively.
The lowest ro-vibrational state of the product K$_2$ molecule lies
$\Delta/(hc)=237$ cm$^{-1}$ beneath that of the reactant KRb molecule
(See Fig.~\ref{fig:rates}a).
Collisions can produce K$_2$ molecules with $v'$ up to $2$ in a multitude of
rotational states $j'$ (up to 63 for $v'=0$; 49 for $v'=1$; 28 for $v'=2$).
We omit coupling of the orbital angular momenta with both the electron and
nuclear spins. Moreover, our  results are restricted to total angular momentum
$J=0$. Fortunately, we can still compare directly with experimental results in
the ultracold regime as only s-wave collisions contribute (that is, only $J=0$
is required for KRb in the ground rotational state $j=0$).
For non-zero $J$, the computational cost scales prohibitively as
$\mathcal{O}((J+1)^3)$ even when we take advantage of parity and exchange
symmetries.

We use the atom-diatom scattering formalism  developed by Pack and
Parker~\cite{pack.parker:quantum,kendrick.pack.ea:hyperspherical}. In the
short-range region we use adiabatically-adjusting-principle-axis hyperspherical
coordinates (APH), an approach which ensures that all atom-diatom arrangements are treated
equivalently, while in the long-range region we use Delves hyperspherical
coordinates (DC) where a molecular basis is more appropriate. For details see the
Methods section and our recent application of this approach to the ultracold
reactive scattering of LiYb molecule with Li atom~\cite{makrides.hazra.ea:ultracold}

In this article, we focus on state-to-state reaction rate coefficients
$K_{v'j'}(E)$ as functions of the collision energy $E$,  $v'$-resolved rate
coefficients $K^{e/o}_{v'}(E)$ obtained by summing over even and odd product
rotational levels $j'$ and, finally, the total rate coefficient $K(E)= (5
K^{e}_{v'}(E)+4K^{o}_{v'}(E))/9$ which accounts for the unresolved nuclear
spin degeneracies of $^{40}$K$_2$ \cite{makrides.hazra.ea:ultracold}.

Figure~\ref{fig:rates}b shows the $J=0$ vibrationally resolved and total
reaction rate coefficients as a function of the collision energy $E$,
along with the experimental total rate coefficient of
Ospelkaus \textit{et al.}~\cite{Science2010}.
Both data for the full trimer potential and that for the pairwise potential
are shown. The differences are seen to be small.
The rate coefficients reach the Wigner threshold regime for $E/k<1$ $\mu$K
and drop off sharply for larger kinetic energies. The drop-off is
consistent with the  unitarity limit, $v_r\pi/k^2_r$, for a single entrance
partial wave between KRb and K. Here, the relative velocity $v_r$ and wavevector
$k_r$ are defined by $E=\mu_r v_r^2/2=\hbar^2k_r^2/(2\mu_r)$ with an atom-dimer
reduced mass $\mu_r$. This rate coefficient is the absolute upper bound to
any ultra-cold reactive process.
The vibrationally resolved rate coefficients have the same functional form as
the total rate coefficient. The most-deeply bound $v'=0$ is most populated
followed by $v'=1$ and $v'=2$.
The experimental result was obtained at a
temperature of 250~nK and is seen to be around 50\% larger than ours.
We note that our calculations do not include all the effects present in the
experiment such as: magnetic field, spin, and conical intersection effects.
We attribute the difference between our calculation and
the experiment to these effects.
For this system the conical intersection is submerged and lies below the
asymptotic energy of the entrance channel, as such the usual vector potential
approach for including the geometric phase (GP) cannot be used as the
singularity at the intersection is exposed.
Therefore in order to include the GP effect the explicit inclusion of the
excited doublet surface is necessary.



Figure~\ref{fig:rates}c shows the thermally-averaged reactive rate
coefficient for the full trimer potential surface as a function of
temperature up to 0.1~K evaluated using the $J$-shifting approach~\cite{Bala2013}.
The rate coefficient is seen to have a minimum near $T=30$ $\mu$K  and slowly
increases for larger temperatures.  For comparison the non-thermalized EQM
results for $J=0$ are repeated and we emphasize that thermalization does not
affect the rate coefficient in the Wigner-threshold regime.

An EQM calculation does not easily lend itself to an intuitive understanding
of the collision. To help explain the total reaction rate coefficient we
have  performed a KRb+K universal-model (UM) calculation
\cite{Kotochigova2010,Julienne2010}, which only relies on the long-range
dispersion coefficient $C_6$ between the molecule and the atom and the
assumption that no flux is returned from short-range separations.  For KRb
in the $v=0$, $j=0$ ro-vibrational state colliding with K this $C_6$ was
previously determined to be $6905E_ha_0^6$~\cite{Kotochigova2010}, where
$E_h$ is the Hartree energy.

Figure~\ref{fig:rates}c shows the UM rate coefficient  as a function of
collision energy assuming only $s$-wave collisions as well as that obtained
after summing over all relative partial waves. The excellent agreement
of the UM with the $J=0$ EQM result and the $J$-shifting method validates the
approximations contained in the model and tells us that this reaction occurs with
unit probability i.e.\ all flux which reaches the short range reacts.

\begin{figure}[b]
\includegraphics[scale=0.33,trim=20 10 0 70,clip]{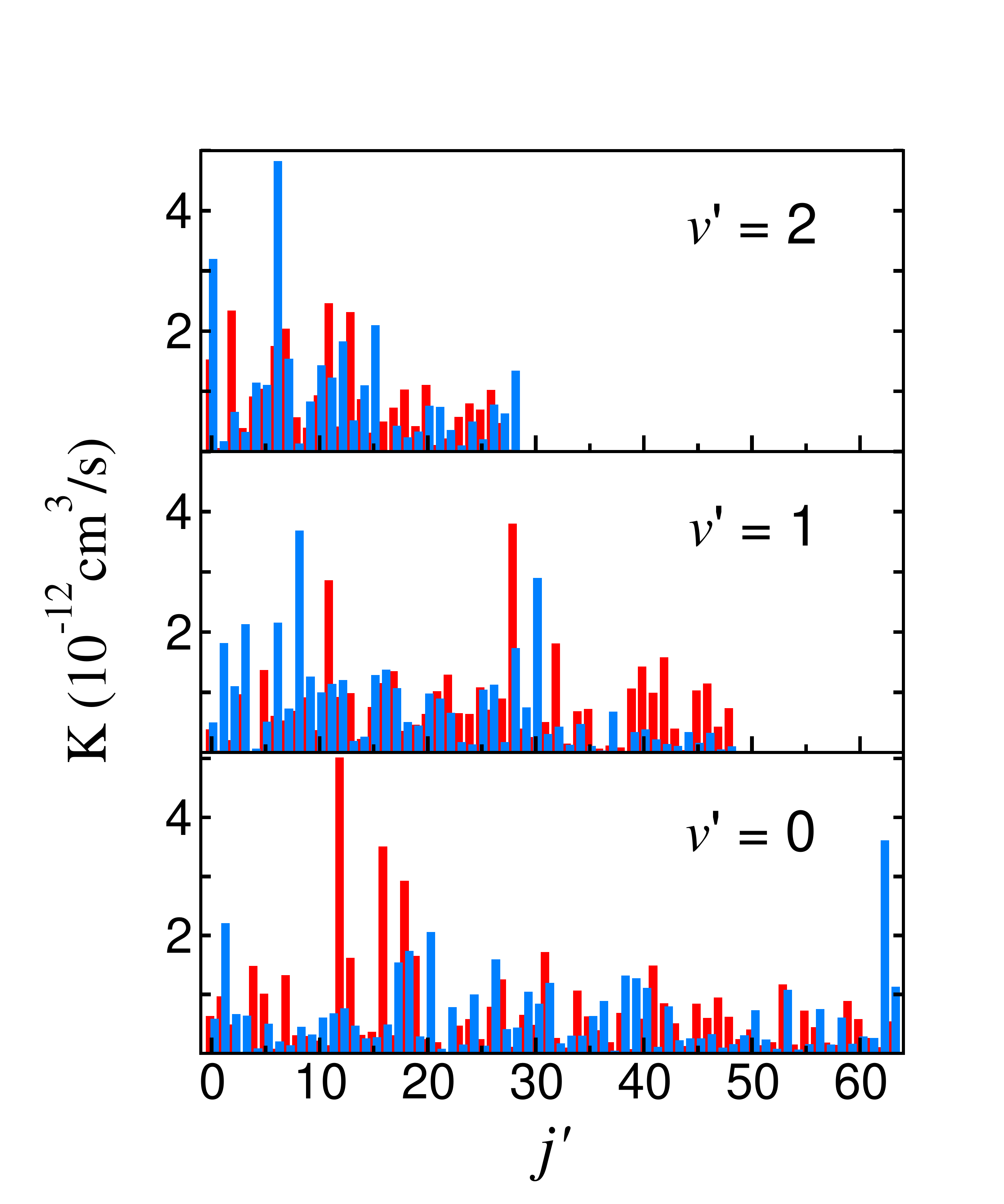}
\caption{{\textbf{$\vert$ Rotationally resolved rates.}} The state-to-state $J=0$ EQM reaction rate coefficients from the
$v=0$, $j=0$ ro-vibrational state of KRb to the $v'=0$, $1$, and $2$ vibrational
level of K$_2$  as a function of product rotational quantum number $j'$.
The blue and red bars are from calculations with the full trimer and additive
pairwise potential, respectively.
Rates are for an initial collision energy of $E/k = 210$~nK.
}
\label{fig:jresolved}
\end{figure}

Controlled chemistry will require the development of experimental techniques
to tune state-to-state rates with an external parameter, such as an
electric or magnetic field.  As a first step in this direction we compute
product $j'$-resolved rates as a function of collision energy.
EQM rate coefficients for $J=0$ at $E/k=210$~nK are shown in
Fig.~\ref{fig:jresolved} for both the full and the pairwise
potential.  The rate coefficients do not vary in the threshold regime and
are thus valid for energies below about 1~$\mu$K. The $j'$-resolved rates
show little obvious structure for either full or pairwise trimer potentials and
are uncorrelated even though the $v'$-resolved and total rate coefficients of
Fig.~\ref{fig:rates}b for these potentials differ only slightly.

\begin{figure}
\includegraphics[scale=0.35,trim=0 10 0 50,clip]{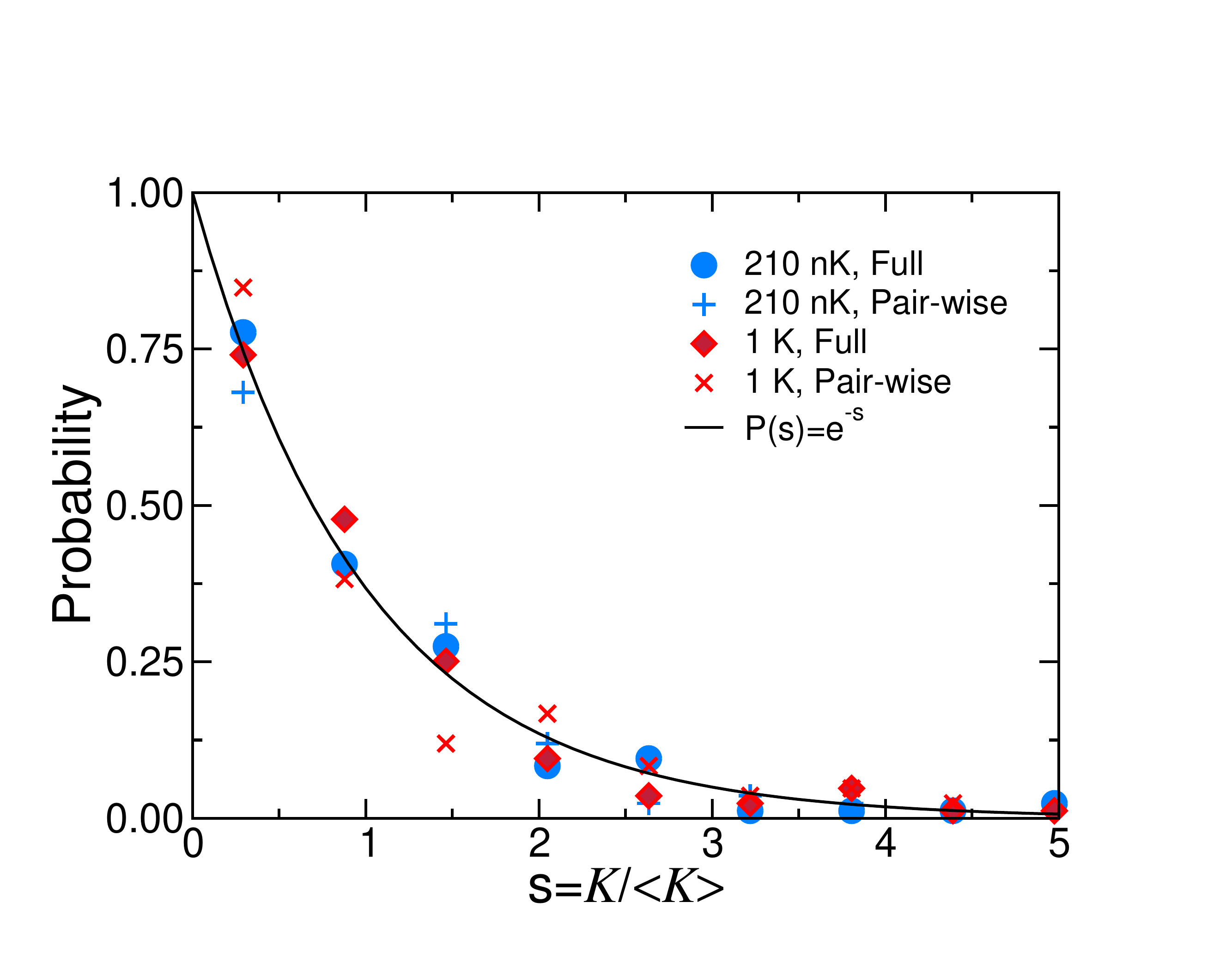}
\caption{{\textbf{$\vert$ Probability distribution of rotationally resolved rates.}} Distribution of the $j'$-resolved rate coefficient.
Blue and red markers correspond to $J=0$ EQM data populating the $v'=0$, 1,
and 2 of K$_2$ at initial collision energy $E/k=210$~nK and 1~K, respectively.
Results for full and pairwise trimer potentials are shown by different markers.
For each collision energy the rate coefficients are scaled to its mean value.
The solid black curve is the Poisson distribution.
}
\label{fig:ROT}
\end{figure}

Given the apparent lack of structure seen in the $j'$-resolved rates the question
is: how much is it possible to say about these rates? The answer is to
think about them statistically.
Fig.~\ref{fig:ROT} presents the data of Fig.~\ref{fig:jresolved} in a different light.
We plot the normalized probability distribution
of the rotationally-resolved rate coefficients $K_{v'j'}(E)$ for two collision
energies and for both the full and pairwise trimer potential.
Each distribution is obtained by binning the $K_{v'j'}(E)$ values into nine
equally-sized bins, up to five times the mean rate coefficient.
Rate coefficients for the three $v'$= 0, 1, and 2 vibrational levels are combined
in order to improve the accuracy of the statistical analysis.
What in Fig.~\ref{fig:jresolved} appeared structureless can now be understood as random
variables sampled from the Poisson distribution.
We note that the Poisson distribution observed
here persists over the entire energy range studied\ (up to 1~K).


\vspace*{2mm}

\noindent
{\bf Chaos  and disorder in  reactive scattering}\\
What is it about reactions of KRb with K which cause this statistical
behaviour? We begin by analyzing the adiabatic potentials in the hyperspherical
coordinate system for various values of hyper radius $\rho$ (computed from
our EQM dynamics calculations).  The appeal of this approach is that it allows
us to visualize and compare the short- and long-range interactions to
understand their effect on the reaction dynamics~\cite{Starace1992,Rahman2001,Daily2015}.

In the asymptotic limit $\rho\to\infty$ the adiabatic potentials converge to
the reactant and product states, whereas for small $\rho$ the shape of the
potentials is defined by the ``variation of the angular potential profile''~\cite{Clary1985}.
For KRbK we include 2543 such curves, for each symmetry, which consists of all channels that are
open at a 220 nK collision energy and those closed channels that are required
to converge the reaction rate. For comparison we perform the same analysis for
the molecular complex \mbox{LiYbLi} formed during the ultracold reaction
LiYb+Li $\to$ Li$_2$+Yb, studied previously by some of us~\cite{makrides.hazra.ea:ultracold}.
The principle difference between these systems is the density-of-states (DOS)
which is much smaller for LiYbLi where only 949 states  for each symmetry are needed for convergence.

\begin{figure*}[t]
\includegraphics[scale=0.23,trim=20 0 0 60,clip]{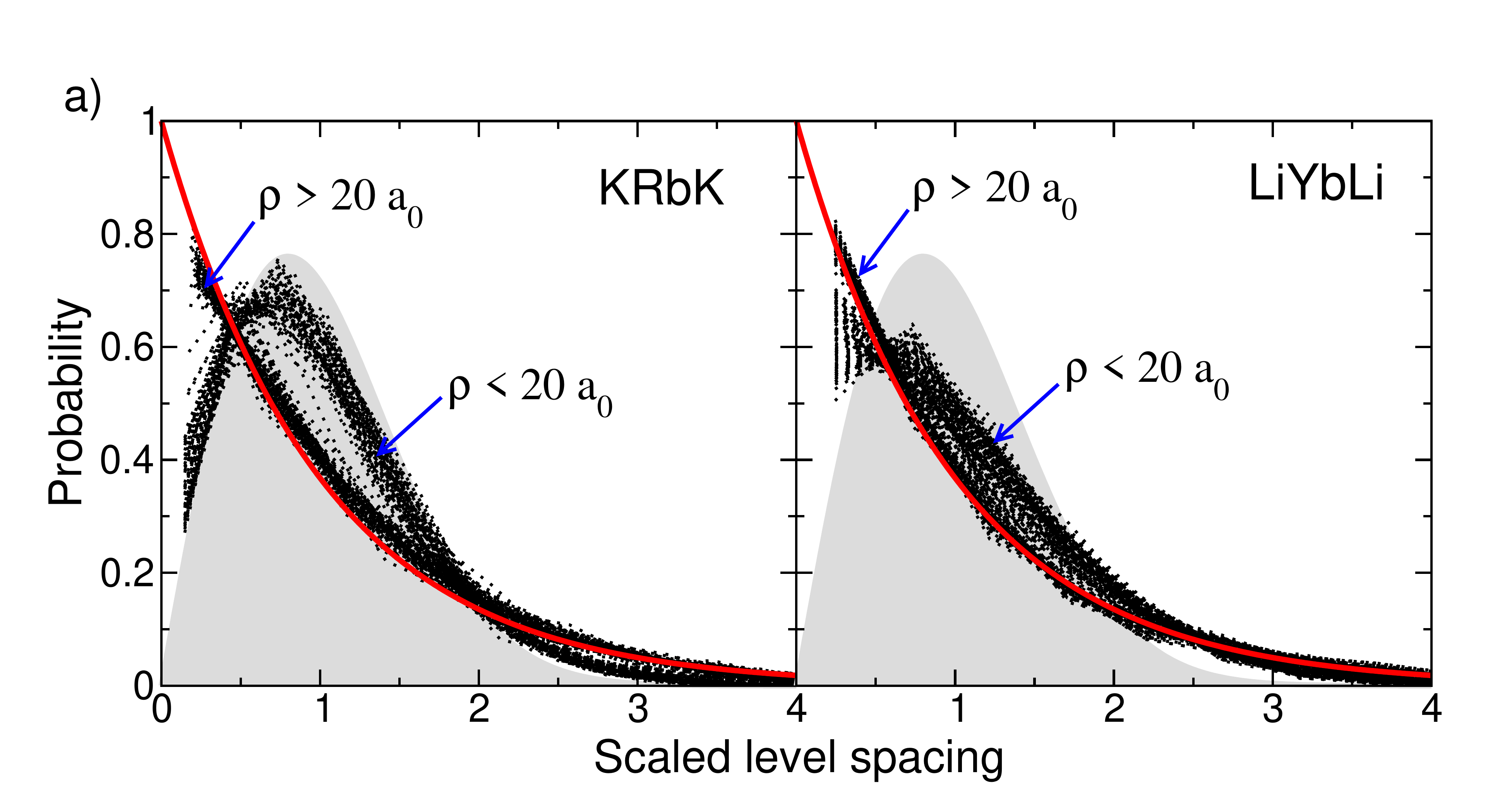}
\includegraphics[scale=0.225,trim=20 0 0 60,clip]{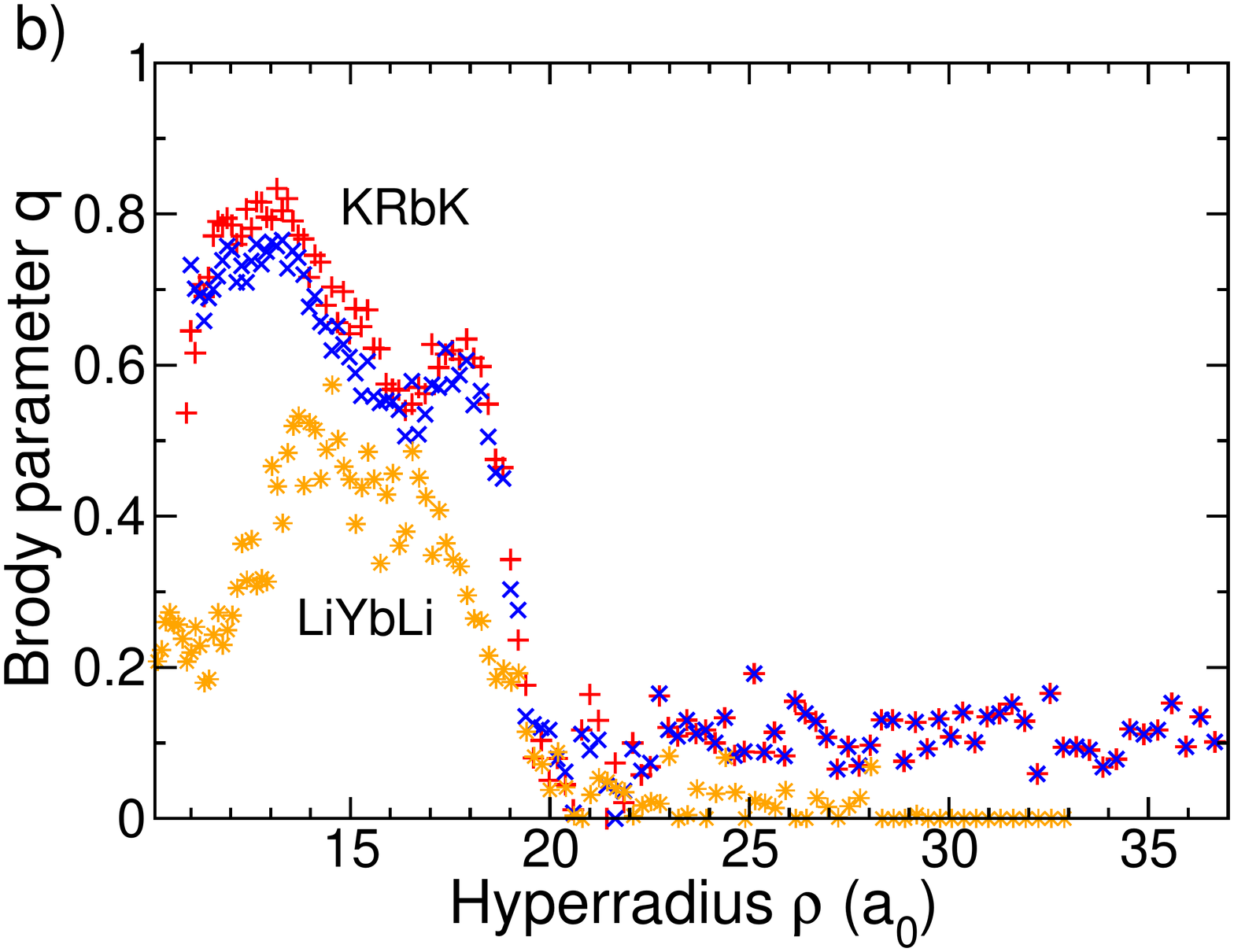}
\caption{{\textbf{$\vert$ Statistical analysis of the short-range adiabatic potentials.}} Panel a: Distribution of nearest-neighbor spacings of $J=0$
KRbK  and LiYbLi  adiabatic potential energies in hyperspherical coordinates.
Each black, dotted curve corresponds to the distribution for a single hyper
radius $\rho$ as a function of scaled level spacing. The shaded grey area and
red curve are Wigner-Dyson and Poisson distributions, respectively.
For both  KRbK  and LiYbLi  we use the full trimer potential.
Panel b: The Brody parameter $q$ as a function of hyper radius for both KRbK
and LiYbLi as derived from the data in panel a for the full trimer potential
and data for KRbK based on the pair-wise potential. Blue and red markers
correspond to distributions for KRbK obtained with the full and pairwise
electronic potentials, respectively, whereas the orange markers for LiYbLi are
obtained  with the full potential. For $q\to 0$ the distribution approaches the
Poisson distribution while for $q\to 1$ it approaches the Wigner-Dyson distribution.
}
\label{fig:RMT}
\end{figure*}

We analyze the adiabatic energy level distribution using random matrix theory
with the goal to reveal collective behavior of complex tri-atomic systems.
Figure~\ref{fig:RMT}a shows nearest-neighbor level-spacing distributions for
KRbK and LiYbLi as a function of scaled spacing between the levels.
Each curve corresponds to the normalized distribution for one hyperradius and
spacings are scaled (divided by the mean spacing).
For KRbK  we can clearly identify two general trends in the distributions
depending on the value of the hyper radius $\rho$. For $\rho<20a_0$, defining
the collisional complex, the adiabatic energy levels exhibit Wigner-Dyson behavior
characteristic of  quantum chaos, due to the strong interactions.
The situation dramatically changes at larger $\rho$, where reactants and
products of the reaction are energetically well separated. Here, the
distribution is described by regular Poisson-like behavior and statistical
disorder.
Figure~\ref{fig:RMT}a also shows the statistics of the adiabatic energy levels
for another trimer complex, LiYbLi. A comparison between the two systems
shows that the chaotic regime in the inner region is more pronounced in
KRbK than for LiYbLi and we attribute this to the much higher DOS of KRbK.

Figure~\ref{fig:RMT}b  presents the Brody parameter $q\in[0,1]$ as a function of
hyperradius for both KRbK and LiYbLi.  It is obtained from fits of
$P(s;q)\propto  s^q\exp[-\alpha(q) s^{1+q}]$ with scaled level spacing
$s$~\cite{Brody1981} to the data in Fig.~\ref{fig:RMT}a. For $q \to 0$ and
1 the distribution approaches the Poisson and Wigner-Dyson distribution respectively.
We see that for both molecules at large $\rho > 20 a_0$, where the reactant
and product molecular levels are well separated and do not repel each other,
the Brody parameter $q$ has a tendency towards the Poisson distribution.
For smaller $\rho$, the Brody parameter becomes much larger for both systems,
although for LiYbLi it is always smaller than that for KRbK,
consistent with the curves in  panel a.
The maximum Brody parameter is $q_{\rm max}=0.85$ for KRbK and 0.55 for LiYbLi.
We identify the chaotic character of the energy levels at short range in KRbK as the
cause of the statistical nature of the $j$-resolved rates discussed earlier.
The large Brody parameter indicates that many strongly interacting
channels contribute to the scattering process. It is the complex interference
between these channels which then leads to the Poisson distribution of $j$-resolved
rates.
The maxima of $q$ versus $\rho$ near $\rho=13a_0$ and $18a_0$ for the KRbK
curves correspond to hyperradii where the energetically-lowest adiabatic
hyperspherical potential has its global and a local minimum, respectively.
Finally, we note that even though the effect of the non-additive three-body
term in the KRbK potential on the distribution is small, it is large enough
to change the product rotational distribution. This is discussed further in
the next subsection.

\vspace*{2mm}

\noindent
{\bf Discussion}\\
We have performed the first numerically-exact quantum-mechanical calculations
for the reactive collisions of a K atom with a KRb molecule in the ultracold
regime. Such calculations are prohibitively expensive, even when neglecting spin and
magnetic field effects, especially for heavy alkali system such as this.
We therefore offer these results as a benchmark for future method development
desperately needed to guide controlled chemistry techniques.

In addition to these calculations we have applied universal quantum-defect theory
to better understand the collisional physics of the reaction forming K$_2$ molecules.
We find that the reactive rate coefficients of both exact and universal
approaches agree  well, suggesting the universal character of the reaction and
confirm predictions of Ref.~\cite{Science2010} for this system.
The agreement between our numerically-exact theory and measurement
of~\cite{Science2010} is good. We attribute the difference to effects not
included in our calculations such as: magnetic field, spin, and conical
intersection effects.

We found that the role of the non-additive three-body contribution to
the potential  and to the total reactive scattering of KRbK is small. This
confirms our initial conclusion that the total reaction rate coefficient does
not depend on details of short-range chemical interactions and is only
defined by the long-range interaction properties. We present this as evidence
for the validity of UM models when applied to  heavy alkali atom+diatom
reactions in general.

We have also presented the first prediction of ro-vibrationally-resolved
reaction rates for this system. We show that $j$-resolved rates can be understood as random
variables sampled from a Poisson distribution. The cause of this statistical
behaviour is the chaotic nature of complex tri-atomic systems which we
quantify with the Brody parameter.
We see that the rotational distributions obtained with the full and
pairwise potentials are completely different, though both still well described by
a Poisson distribution. As this distribution is merely a consequence of the chaotic nature
of the complex at short hyperradius this is not specific to reactions
of K with KRb. We predict that all ultracold atom+dimer and dimer+dimer reactions
involving heavy alkali atoms will exhibit this same Poisson distribution of
$j$-resolved rates, provided there are sufficient product channels to perform a
statistical analysis.

Our future plans involve a full treatment of the conical intersection and
accompanying geometric phases (GP), by including the excited doublet state in
the  calculations.  The GP effect on chemical reactions is well studied,
but is often masked by thermal averaging~\cite{kendrick:geometric*2}.
There is no thermal averaging in the ultracold domain, where the GP has been
predicted to have a significant effect~\cite{kendrick.hazra.ea:geometric,hazra.kendrick.ea:importance}.
We also plan to investigate if the system exhibits a quantum butterfly effect
where tiny perturbations in the interaction potential lead to exponentially
different results (suggested by the completely different $j$-resolved rates
seen when the three body term in the potential is included or not).
The presence of chaos in such systems would have important implications for
the development of controlled chemistry.
Studies have shown that small carefully-chosen external perturbations can
produce a large beneficial change in the system
behavior~\cite{Garfinkel1992,Petrov1993,Shinbrot1993}.

We have shown that ultracold reactions of heavy alkali systems are inherently
statistical in nature. We believe that this has
significant implications for the development of controlled chemistry
as well as suggesting a way forward in attacking the currently
intractable computational challenge such systems pose.

\vspace*{4mm}

\noindent
{\bf Methods}\\

{\bf Non-additive full  trimer potential.}
\/We have computed the ground-state three-dimensional surface $U({\vec r})$
for KRbK using the multi-reference configuration-interaction (MRCI) method
within the MOLPRO software package, where
$\vec r=(r_{\rm K(1)Rb}, r_{\rm K(2)Rb}, r_{\rm K(1) K(2)} )$
and $r_{{\rm K}(i){\rm Rb}}$ are the separations between the $i$-th K atom and
Rb and $r_{\rm K(1) K(2)}$ is the separation between the two K atoms.
For the closed shell electrons of   K and Rb  we employed the
ECP18SDF and ECP36SDF energy-consistent, single-valence-electron,
relativistic pseudo-potentials of the Stuttgart/Cologne groups
\cite{Stoll1983,Stoll1985}.  Core polarization potentials (CPPs) are
modeled after Ref.~\cite{Meyer1984} with cutoff functions with exponents 0.265 and
0.36 for Rb and K, respectively.  So each atom is described by a single
electron model. We used an uncontracted {\it sp}\/ basis set supplemented
with ECP core potentials and augmented by additional s, p, d, and f functions~\cite{JHutson2010}.

The {\it ab~initio} electronic structure computations are too
expensive to be used for each geometry of the collisional complex.
Consequently, following~\cite{Aguado1992} our discrete data on the three-body contribution
(with the pairwise contribution removed) is fitted to a
suitable analytical functional form, where adjustable parameters  are found
with a least-squares procedure. We find that the
fit reproduces the PES without introducing spurious features.

\begin{figure*}
\includegraphics[scale=0.30,trim=0 0 0 0,clip]{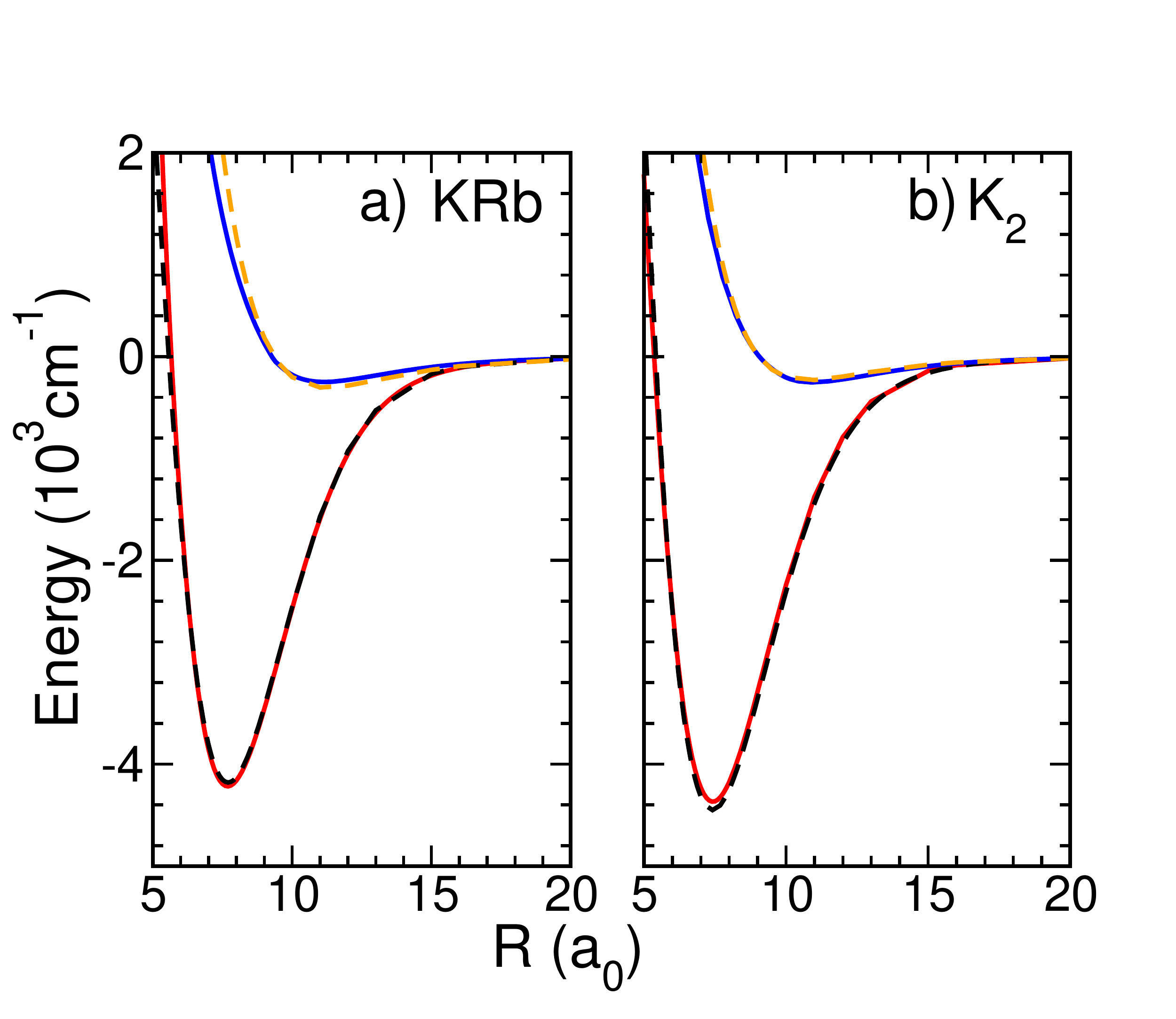}
\includegraphics[scale=0.30,trim=0 0 0 0,clip]{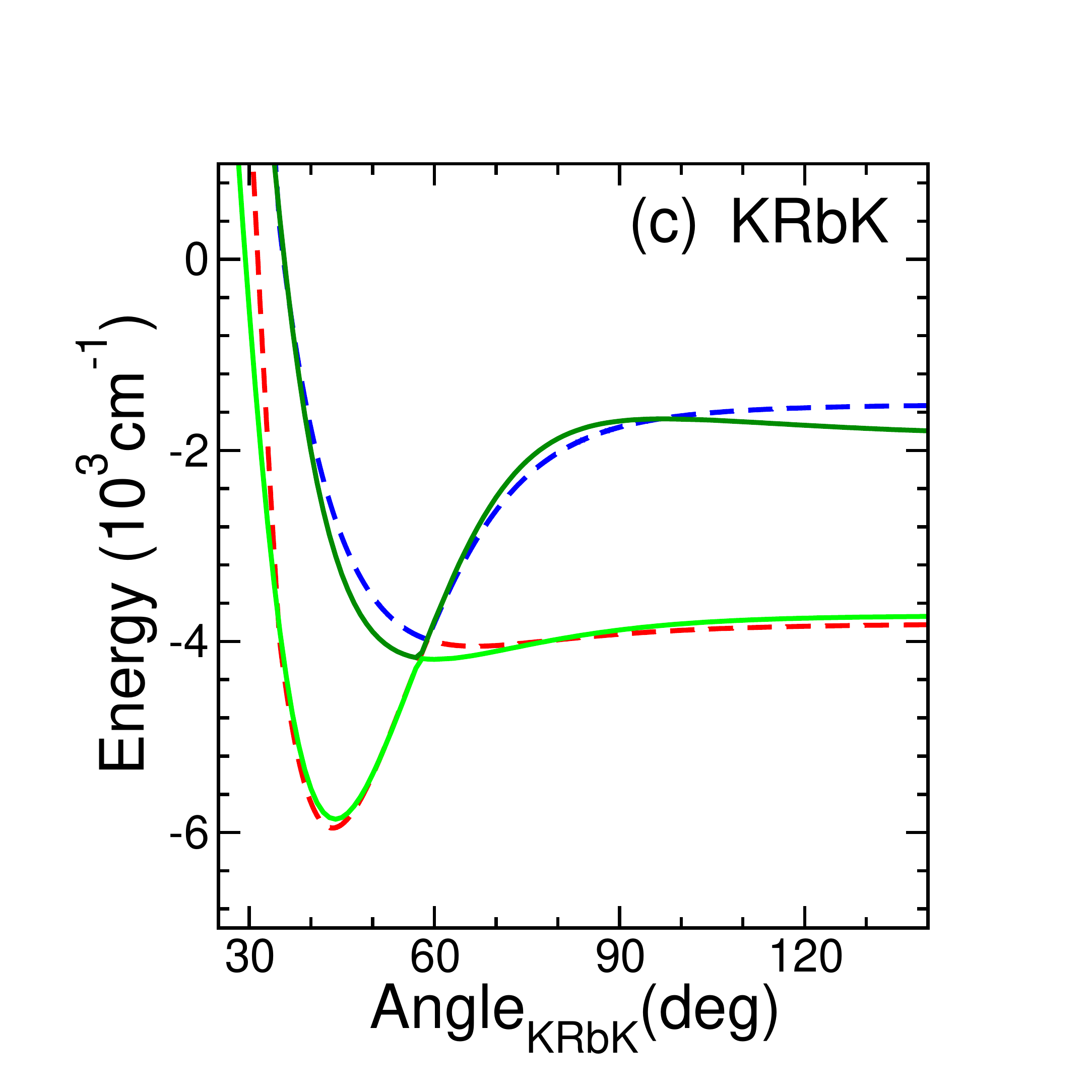}
\caption{{\textbf{$\vert$ Dimer and trimer potential energy curves.}} The calculated singlet X$^1\Sigma^+$ and triplet a$^3\Sigma^+$ potentials
of KRb (solid lines in panel a) and K$_2$ (solid lines in panel b) as a
function of inter-atomic separation using the same basis set and CPP as for the KRbK trimer.
The dashed lines in both panels are the corresponding spectroscopically-accurate
dimer potentials~\cite{Tiemann2007,Tiemann2008}.
Panel c: A one-dimensional cut through the two energetically-lowest adiabatic
KRbK trimer potential surfaces  based on full (solid lines) and pairwise (dashed lines)
calculations for $R_{\rm K(1)Rb}=R_{\rm K(2)Rb}=10a_0$.
}
\label{fig:test}
\end{figure*}

We also computed the {\it ab~initio} dimer X$^1\Sigma^+$ and a$^3\Sigma^+$
potentials for KRb and K$_2$ at the same level of electronic-structure theory
as the trimer potential. These dimer potentials are shown in
Figs.~\ref{fig:test}a and b and compared with spectroscopically-accurate
potentials of KRb \cite{Tiemann2007} and  K$_2$ \cite{Tiemann2008}. Agreement
is better than 38 cm$^{-1}$ for KRb and 83 cm$^{-1}$ for K$_2$ at the equilibrium separation.
The comparison of these potential curves provides evidence that the
non-additive term in KRbK is small.

\begin{figure*}
\includegraphics[scale=0.35,trim=0 0 0 20,clip]{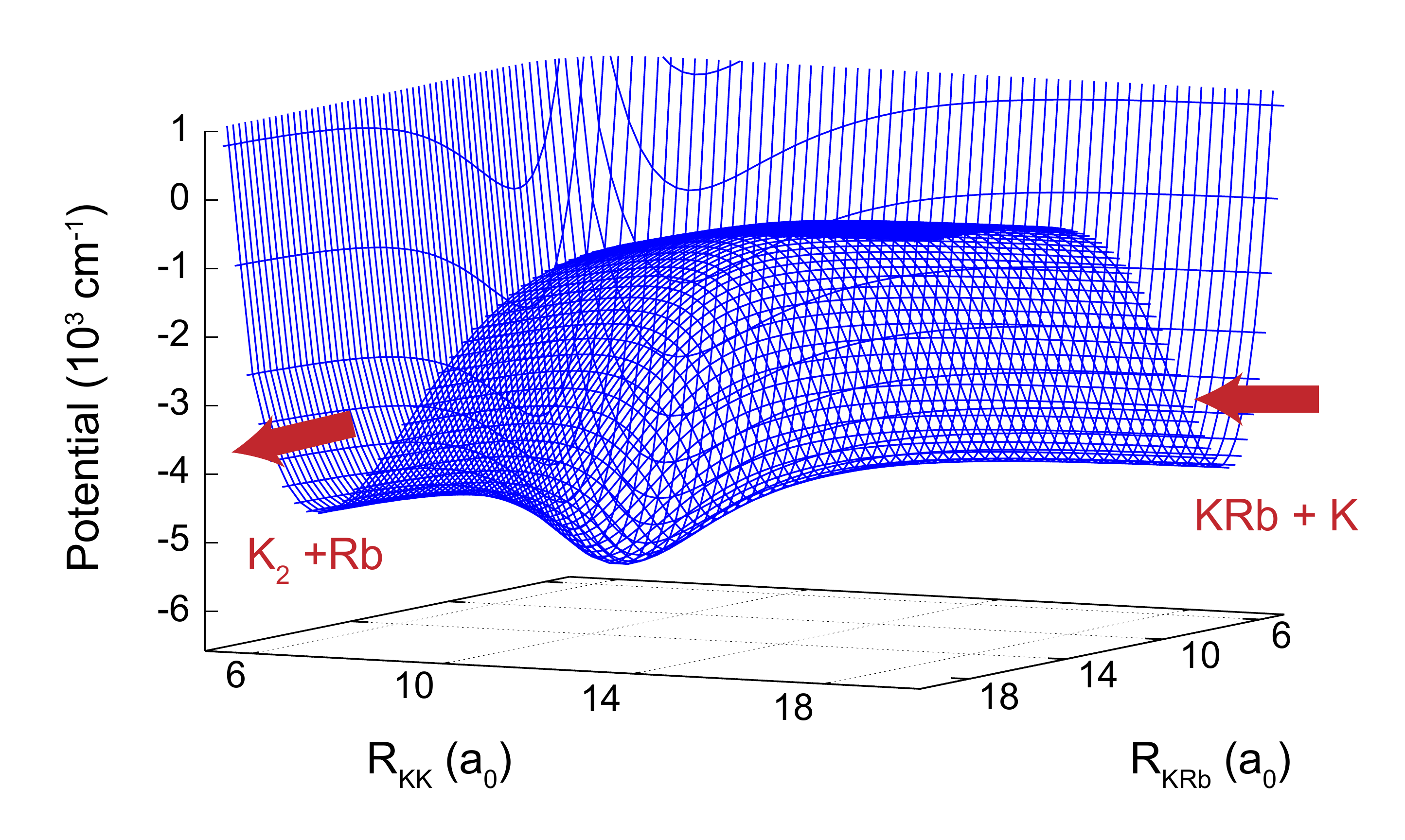}
\caption{{\textbf{$\vert$ K-KRb potential energy surface for collinear geometry.}} A two-dimensional cut through the energetically-lowest
$^2A'$ adiabatic potential energy surface of the KRbK trimer as a function of the K$-$Rb and K$-$K bond lengths along the collinear geometry.
The zero of energy corresponds to the energy of three well-separated atoms. The arrows indicate the entrance and exit channels of the chemical reaction.
}
\label{KRbK_linear}
\end{figure*}

Figure~\ref{KRbK_linear} shows a two-dimensional cut through the energetically-lowest
$^2A'$ adiabatic potential energy surface of the KRbK trimer as a function of the K$-$Rb and K$-$K bond lengths
for the collinear geometry. This figure clearly shows that the potential of the collisional complex, where all atoms are
close together, is deeper than that of the reactant and product configurations with atoms spending some time moving
within the complex before reacting and flying out as K$_2$ + Rb.


\noindent
{\bf Pairwise trimer potential.}
For our coupled-channels calculation we have also used a doublet trimer potential constructed from the pairwise
singlet $V^X_{ij}(r)$ and triplet $V^a_{ij}(r)$ potentials. Here indices $i$ and $j$ label the atoms and $r$ is their separation.
The wavefunction of ground-state alkali-metal atoms is uniquely described by its electron spin $s_i=1/2$,
where $i$=A, B, or C.  Two such atoms couple to spin states $|S_{ij}\rangle \equiv|(s_is_j)S_{ij}\rangle$,
with a total spin singlet $|0\rangle =|(\frac{1}{2} \frac{1}{2}) 0\rangle$ and triplet
$|1\rangle =|(\frac{1}{2} \frac{1}{2}) 1\rangle$ state. Their pairwise potential Hamiltonian is denoted by
$\hat V_{ij}(r) = V^X_{ij}(r) |0\rangle\langle 0|+V^a_{ij}(r)|1\rangle\langle 1|$.
Three  atoms couple to states $|(s_{\rm A}s_{\rm B})S_{\rm AB}, s_{\rm C};S\rangle$.  For  a doublet total  electron spin of $S=1/2$ there exist two ways to couple
the spins. These are $|-\rangle_{\rm C}= |(\frac{1}{2} \frac{1}{2}) 0, \frac{1}{2}; \frac{1}{2}\rangle$ and $|+\rangle_{\rm C}=|(\frac{1}{2} \frac{1}{2}) 1, \frac{1}{2};\frac{1}{2}\rangle$, where  subscript C in $|\pm\rangle_C$ indicates that first the spins of atoms A and B are coupled together to a total spin $S_{\rm AB}$, which is then coupled to that of atom C.
The pairwise trimer potential  $\hat V_{\rm AB}(r_{\rm AB})+\hat V_{\rm BC}(r_{\rm BC})+\hat V_{\rm CA}(r_{\rm CA})$
 leads to a $2\times2$  matrix in  the $|\pm\rangle_C$ basis. It is most-easily constructed by using angular momentum algebra
to transfer between the equivalent basis sets $|\pm\rangle_A$, $|\pm\rangle_B$, and $|\pm\rangle_C$.
Diagonalization of this Hamiltonian matrix leads to the two adiabatic $^2$A potentials
\begin{widetext}
\begin{equation}
U^{(\pm)}_{\rm pw}(\vec r) 
 =V^d_{\rm AB}(r_{\rm AB}) + V^d_{\rm BC}(r_{\rm BC} )+ V^d_{\rm CA}(r_{\rm CA} )
        \pm  \sqrt{
       \left( \begin{array}{l}
              (V^{e}_{\rm AB}(r_{\rm AB}))^{2} \\\quad + (V^{e}_{\rm BC}(r_{\rm BC}))^{2} \\
              \quad\quad+ (V^{e}_{\rm CA}(r_{\rm CA}))^{2}
              \end{array} \right )
      -
\left(  \begin{array}{l}
       V^{e}_{\rm AB}(r_{\rm AB}) V^{e}_{\rm BC}(r_{\rm BC}) \\
       \quad + V^{e}_{\rm BC}(r_{\rm BC}) V^{e}_{\rm CA}(r_{\rm CA})  \\
       \quad\quad+ V^{e}_{\rm CA}(r_{\rm CA} ) V^{e}_{\rm AB}(r_{\rm AB})
  \end{array}  \right) }  ,
  \end{equation}
\end{widetext}
where the dispersion ({\it d}) and exchange ({\it e}) potential are
\[
     V^{d/e}_{ij}(r) = (V^X_{ij}(r) \pm V^a_{ij}(r))/2 \,,
\]
respectively. For our KRbK trimer A=K(1), B=Rb and C=K(2).
By construction the factor in the square root is non-negative. Hence, potential $U^{(-)}_{\rm pw}(\vec r)$
has the lowest energy and is used in our exact quantum-mechanical (EQM) calculations. As an aside we note that in the $C_{2v}$ symmetry  the factor in the square root can be zero, the two potentials are degenerate
and the system has conical intersections.

\noindent
{ \bf Exact Quantum Dynamics Calculations.}
\/For details of the EQM method see our previous
paper~\cite{makrides.hazra.ea:ultracold} on LiYbLi and references therein,
which also contains details of the calculation of the LiYbLi adiabatic
potential energies used in this work. Here we outline details specific to the
KRbK calculation.

The EQM calculations can broadly be split into three main steps: the
numerical computation of 5D hyperspherical surface functions in the APH
coordinates in the short-range region and DC in the long-range region; the
log-derivative propagation of the CC equation in these coordinates; finally,
the asymptotic matching to ro-vibrational states in Jacobi
coordinates.

The 5D APH surface functions in the short-range region, from $\rho=8.0a_0$ to $38.15a_0$,
are functions of two internal coordinates $\theta$ and $\phi$ and three Euler
angles $\alpha$, $\beta$, and $\gamma$ to orient the molecule in space.
APH surface functions in $\theta$ and $\phi$ are determined by  $l_{\rm max}$
and $m_{\rm max}$ respectively. This region is further subdivided into six,
with increasing $l_{\rm max}$ and $m_{\rm max}$ to ensure converged surface functions.
These regions are $8.0 a_0<\rho < 11.33 a_0$, $11.33 a_0<\rho< 16.87 a_0$,
$16.87 a_0<\rho < 18.82 a_0$, $18.82 a_0<\rho< 22.74 a_0$,
and $22.74 a_0<\rho < 32.21 a_0$
with $l_{\rm max}=103,117,123,149,169,189$ and $m_{\rm max}=206,234,246,298,338,378$, respectively.
For $J=0$ this leads to  5D surface function matrices of dimension
42\,952, 55\,342, 61\,132, 89\,550, 115\,090, and 143\,830. Explicit diagonalization of such large matrices
is not computationally tractable so we use the sequential
diagonalization truncation technique~\cite{bacic.whitnell.ea:localized,bacic.kress.ea:quantum} to reduce the dimension and the
implicitly restarted lanczos method~\cite{sorensen:implicit} to compute only the lowest $2543$ surface functions
of each exchange symmetry needed for the log-derivative propagation.
These surface functions are computed on a logarithmic grid in $\rho$ with
157 sectors.

Delves coordinates are used in the outer region from $\rho=32.21a_0$ to
$\rho_{\rm max}=174.95a_0$ where a linear grid in $\rho$ is used with 456 sectors.
The number of basis functions is determined by an energy cutoff of 0.3~eV
relative to the minimum energy of the asymptotic K$_2$ diatomic potential.
A one-dimensional Numerov method is used to compute the adiabatic surface functions.

The log-derivative matrix is propagated separately for each exchange
symmetry and parity, though only even parity is needed in this work as $J=0$.
In the APH region the propagation includes 2543 channels of each
symmetry, over 20\% of which are closed at all $\rho$.
The DC region only requires 423 for each symmetry as
many of the channels locally open at short range have become strongly closed.
Consequently the long-range propagation takes a negligible time compared to
the short-range as the computational cost scales as the number of
channels cubed.

At $\rho=\rho_{\rm max}$, we match the DC wave functions to
asymptotic channel functions corresponding to ro-vibrational
levels of the KRb and K$_2$ molecules, defined in Jacobi coordinates.
This includes vibrational levels up to 2 for KRb and 5 for K$_2$ with
rotational levels up to a maximum of 68 and 93 respectively.

\section*{Data availability}
The data that support the findings
of this study are available from the authors on reasonable request,
see author contributions for specific data sets.

\section*{Acknowledgments}
We acknowledge support from the  US Army Research Office, MURI grants No.~W911NF-12-1-0476 and
W911NF-14-1-0378, the US National Science Foundation, grants No.~PHY-1505557, PHY-1619788, and
PHY-1125915.
In addition, SK acknowledges support from the US Air Force Office of Scientific
Research, grant No. FA9550-14-1-0321 and BKK acknowledges that part of this
work was done under the auspices of the US Department of Energy,
Project No. 20170221ER of the Laboratory Directed Research and Development
Program at Los Alamos National Laboratory.
Los Alamos National Laboratory is operated by Los Alamos National Security,
LLC, for the National Security Administration of the US Department of Energy
under contract DE-AC52-06NA25396.

\section*{Author Contributions}
N.B.\@ and S.K.\@ conceived the research project. S.K.\@, C.M.\@, M.L.\@, and A. P.\@ contributed to the computation of the PES,
statistical analysis of quantum dynamics results,
and application of the universal model.
J.F.E.C.\@ carried out the quantum dynamics calculations with assistance
from N.B.\@ and B.K.K.\@ J.F.E.C.\@ and S.K.\@ wrote the manuscript and all
authors reviewed and commented on the manuscript.

\subsection*{Additional Information}

\textbf{Competing financial interests:}\ \ The authors declare no competing financial interests.


\end{document}